\documentstyle[twoside,fleqn,espcrc2]{article}

\newcommand{\AmS}{{\protect\the\textfont2
  A\kern-.1667em\lower.5ex\hbox{M}\kern-.125emS}}
\title{Particle with torsion  on $3d$ null-curves }
\author{
A. Nersessian\address{Laboratory of Theoretical Physics,
Joint Institute
for Nuclear Research,
Dubna, 141980, Russia}
\address{Department of  Physics, Yerevan State University,
A.Manoukian St., 3, Yerevan 375049, Armenia}
\thanks{Partially supported by  INTAS-96-538
and HLP-99-10 grants},
R. Manvelyan\address{
 Department of Physics, University of Kaiserslautern, P.\ O.\ Box 3049,
 Kaiserslautern, D67653 Germany}\thanks{Alexander von Humboldt fellow,
on leave from Yerevan Physics Institute, Yerevan, Armenia}
 and H.J.W. M\"uller-Kirsten$^{\rm c}$ }
\begin{document}
\begin{abstract}
We consider a $(2+1)$-dimensional  mechanical system
 with the Lagrangian linear in the torsion of a light-like curve.
We give  Hamiltonian formulation of this system and
show that its mass and spin spectra are defined
by one-dimensional nonrelativistic mechanics with a cubic potential.
Consequently, this system possesses the properties
typical of resonance-like particles.
\end{abstract}
\maketitle
\section{Introduction}
 The search of Lagrangians, describing spinning particles, both
 massive and massless, has a long story.
 The conventional  approach in this direction consists
 in the extension of the initial space-time
 by  auxiliary odd/even coordinates which equip a system  with
 spinning degrees of freedom.
 There is another, less developed approach,  where the spinning particle
 systems are described by the Lagrangians,
 which are formulated in the initial space-time, but
 depend on higher derivatives.
 The aesthetically attractive point of the latter approach is
 that spinning degrees of freedom are encoded in the geometry
 of  trajectories.
The Poincar\'e and reparametrization invariance
require    actions to be of the form
\begin{equation}
 {\cal S} =\int {\cal L}({ k}_1,....,{ k}_N)|d{\bf x}|, \quad N\leq D-1,
\label{gactions}\end{equation}
 where $k_I$  denote the reparametrization invariants
 (extrinsic curvatures)  of  curves,
\begin{equation}
{ k}_I
=\frac{\sqrt{\det{\hat g}_{I+1}\det{\hat g}_{I-1}}}{\det{\hat g}_I },
\end{equation}
where
${(g_I)}_{ij}
\equiv
{\partial_{(i)}{\bf x}}{\partial_{(j)}{\bf x}}$, $i,j=1,\ldots,I.$

It was shown by M.Plyushchay that  a four-dimensional  system of this sort
with Lagrangian ${\cal L}=ck_1$ describes  a massless particle  with the
helicity $c$ \cite{massless}. This model has
$W_3$ gauge symmetry \cite{rr}: its  classical trajectories are
  {\it space-like plane curves} with arbitrary first curvature.
Higher-dimensional generalization of this model is given by the
action \cite{qfthep}
\begin{equation}
{\cal S}=c \int{k}_N|d{\bf x}|, \quad N\leq [(D-2)/2].
\label{action}\end{equation}
This system has the following  interesting properties:
\begin{itemize}
\item This is the only model that  leads to an irreducible
 representation of the Poincar\'e group. It describes  massless particles
specified by the coinciding  weights of  $SO(N)$ group.
\item It has $N+1$ gauge degrees of freedom:
 the classical solution of this model is a  space-like
     curve spe\-ci\-fied by the relations:
     $k_1,\ldots k_N$ are arbitrary;
    $k_{N+a}=k_{N-a},\;k_{2N}=0$, $a=1,\ldots, N-1$
($W_{N+2}$ gauge symmetry? \cite{rr1}).
\end{itemize}

All massive  models with an  action (\ref{gactions})
correspond to reducible representations of the Poincar\'e group.
Nevertheless, these models can be useful in planar physics,
where a value of spin can be arbitrary.
Extensive studies in this direction were inspired  by the remarkable
work of Polyakov \cite{polakov}
 where the  $CP^1$ model with the Chern-Simons term was investigated.
Evaluating the effective action for the charged solitonic excitation
he found that it is of the type (\ref{gactions}), where
${\cal L}=m_0 +\frac{c^2}{m_0}k_2$.
Later it was shown \cite{misha}
that though the  trajectories of the system are
 {\it time-like}, it has  not only massive, but also
 tachionic and  massless sectors, with mass  and spin related by
the Majorana condition
    \begin{equation}
{\rm Spin}\times{\rm Mass }=c^2,
\label{majorana}\end{equation}
while  $|{\rm Mass}|\geq m_0$.

Adding,  to the initial Lagrangian, the term proportional to $k_1$
modifies the spin-mass relation but preserves the basic properties
 of the initial model \cite{misha2}.

In a $(2+1)$-dimensional space one can
 consider  actions of another sort
 defined on  the  light-like (or null) curves
\begin{equation}
{\cal S}=\int {\cal L}({K})d{\sigma},\quad
d\sigma=|{\ddot{\bf x}}|^{1/2}du,\quad {\dot{\bf x}}^2=0,
\end{equation}
where  $K=|d^3{\bf x}/d\sigma^3|^2$ is the  torsion for light-like curves.

In Ref. \cite{nr},  the simplest system of this sort was considered
given by the action
  \begin{equation}
{\cal S}=2c\int d\sigma .
\label{action0}\end{equation}
It was found that it describes the anyons with Majorana-like spectrum
(\ref{majorana}), while its classical solutions are null-helices.

In the present  paper, we consider a more complicated  three-dimensional system,
 associated with null-curves
  \begin{equation}
{\cal S} =2c\int(\epsilon +K)d\sigma .
\label{action1}\end{equation}
 where $\epsilon $  is a  constant, and $K$ is the torsion of a null-curve.

We  show that this system has a much reacher structure
than the previous one and  is related with  nonrelativistic mechanics
$$
d\pi\wedge dq, \quad \pi^2+ q^3-2\epsilon q^2+\frac{ms}{c^2}q+
\frac{m^2}{c^2}=0,
$$
where $m$ and $s$ denote the mass and spin of the system,
while $q=\epsilon-K/c$.

We conclude the Introduction with some basic facts from the geometry of
three-dimensional null-curves to be used in this paper.

For the description of null curves it is convenient to use
the moving frame $({\bf e}_{\pm}, {\bf e}_1)$:
 \begin{equation}
\begin{array}{c}
{\bf e}_{\pm}{\bf e}_{\pm}={\bf e}_\pm{\bf e}_1 =0,  \quad
{\bf e}_+{\bf e}_-=-{\bf e}^2_1 =1,
\end{array}
\label{mf}\end{equation}
with the vector product $\times$ defined as follows
\begin{equation}
  {\bf e}_+\times {\bf e}_-={\bf e}_1,\quad
{\bf e}_\pm\times {\bf e}_1=\pm{\bf e}_\pm.
\end{equation}
In this notation pseudoarch-length $d\sigma\equiv{\tilde\sigma}du$
 and the torsion  $K$ are defined via the Frenet  equations \cite{bonnor}:
\begin{equation}
\begin{array}{c}
{\bf x}'={\bf e}_+,\quad{\bf e}_+'={\bf e}_1, \\
{\bf e}_1'= K{\bf e}_+ +{\bf e}_-,\quad
{\bf e}_-'= K{\bf e}_1 ,
\end{array}
\label{ff}\end{equation}
where $\;'\equiv d/d\sigma$ .\\
Hence,
\begin{equation}
{\tilde\sigma}=-{\dot{\bf e}}_+{\bf e}_1,
\quad 2K={{\bf e}'_1}^2.
\label{pa}\end{equation}

\section{Hamiltonian formulation}
Prior giving the Hamiltonian formulation of the system (\ref{action1}),
let us present, for completeness,
the Hamiltonian system describing (\ref{action0}) \cite{nr}.

The  system (\ref{action0}) is described
 by the  Hamiltonian structure
\begin{equation}
\begin{array}{c}
\omega=d{\bf p}\wedge d{\bf x}+cd{\bf e}_+\wedge d{\bf e}_1 ,\\
{\cal H}=\frac{{\tilde\sigma}}{2c}
\left[c^2{\bf e}_1^2 +({\bf p}{\bf e}_+-2c)^2
+{\bf p}^2{\bf e}_+^2\right]
\end{array}
\end{equation}
and the  constraints
\begin{equation}
\left\{
\begin{array}{c}
{\bf e}_1^2 +1\approx 0,\quad
{\bf e}_+^2\approx 0,\quad {\bf e}_1{\bf e}_+\approx 0,\\
{\bf p}{\bf e}_+ -c\approx 0,\quad
{\bf p}{\bf e}_1\approx 0.\\
\end{array}
\right.
\label{sc}
\end{equation}
Its Lorentz generator is of the form
\begin{equation}
 {\bf J}= {\bf p}\times {\bf x} + c{\bf e}_+ ,
\end{equation}
from which the relation (\ref{majorana}) follows immediately.\\
Introducing
\begin{equation}
 K=-{\bf p}^2/2c^2,\quad  {\bf e}_-={\bf p}/c+K{\bf e}_+,
\end{equation}
we reduce the equations of motion to the Frenet formulae, while
the effective coordinate reads
$${\bf X}\equiv {\bf x} -\frac{c}{{\bf p}^2}{\bf p}_+\;:\;\;
\ddot{\bf X}=0.$$
{\it Thus, massive (tachionic) solutions of (\ref{action0})
correspond to the light-like helices
with negative (positive)  torsion}.

Now let us  give the Hamiltonian formulation  of (\ref{action1}).
Taking into account the Frenet equations (\ref{ff}),
we can replace the initial Lagrangian depending on third derivatives
 by the classically equivalent one depending on first derivatives only
\begin{equation}
\begin{array}{c}
 L={\tilde\sigma}\left[
2c\epsilon+ c{\bf e'}^2_1
+{\bf p}({\bf x'}-{\bf e}_+)+
\right.
\\
\left.
+{\bf p}_+({\bf e'}_+ - {\bf e}_1)
-\sum_{i,j}d_{ij}
({\bf e}_i{\bf e}_j -\eta_{ij})
\right],
\end{array}
\label{lk1}\end{equation}
where
 ${\bf x}, {\bf p}, {\bf p}_{+},{\bf e}_{i}, d_{ij}, {\tilde\sigma}$
are independent variables, $i,j=+,1$;
 $\eta_{++}= \eta_{1+}=0,\eta_{11}=-1$.

The momentum conjugate to ${\bf e}_1$ reads
$${\bf p}_1=\frac{\partial L}{\partial{\bf\dot e}_1 }=
2c{\bf e'}_1.$$
Thus, we get  the primary constraints
$$
{\bf p}_1{\bf e}_+ -2c\approx 0,\quad {\bf p}_1{\bf e}_1 \approx 0 .
$$
Performing the Legendre transformation, after some work, we obtain
 the following Hamiltonian system:
\begin{equation}
\begin{array}{c}
\omega=d{\bf p}\wedge d{\bf x}+d{\bf p}_+\wedge d{\bf e_+}+
d{\bf p}_1\wedge d{\bf e}_1\\
{\cal H}={\tilde\sigma}\left[\phi_0 +
\lambda\phi_1+\sum_{i,j=+,1}d_{ij}u_{ij}\right]
\end{array}
\label{1}\end{equation}
with  constraints
\begin{equation}\left\{
\begin{array}{c}
\phi_0={\bf p_1}^2/4c+{\bf p}{\bf e}_+
+{\bf p_+}{\bf e}_1-2c\epsilon\approx 0;\\
\phi_1={\bf p}_1{\bf e}_+-2c\approx 0,\\
\phi_2={\bf p}_1{\bf e}_1\approx 0,\\
\phi_3={\bf p}_+{\bf e}_+\approx 0,\\
u_{ij}={\bf e}_i{\bf e}_j-\eta_{ij},
\end{array}\right.
\end{equation}
and the
expessions  for the Lagrangian multipliers:
\begin{equation}
2d_{11}={\bf p}{\bf e}_+-{\bf p}^2_1/2c,\quad
2c\lambda={\bf p}_+{\bf e}_1-{\bf p}{\bf e}_+.
\end{equation}
Let us introduce
\begin{equation}
 {\bf e}_-\equiv \frac{{\bf p}_1}{2c}+\frac{1}{2c}(
{\bf p}{\bf e}_+ +{\bf p_+}{\bf e}_1-2c\epsilon){\bf e}_+,
\label{2}\end{equation}
which forms, together with ${\bf e}_+, {\bf e}_1$,
the moving frame.

Thus, ${\bf p}$ and ${\bf p}_+$ are decomposed as follows
\begin{eqnarray}
&{\bf p}_+= y_+{\bf e}_+ - y_1{\bf e}_1,
\label{deco}\\
&{\bf p}/c=q{\bf e}_- +
x{\bf e}_+-\pi{\bf e}_1,&\label{p}
\label{pdeco}\end{eqnarray}
The  equations of motion  for ${\bf x}, {\bf e}_{\pm}, {\bf e}_1$
coincide with  (\ref{ff}),
if we identify
\begin{equation}
   K=\epsilon-{\bf p}{\bf e}_+/c\equiv\epsilon - q,
\label{K}
\end{equation}
while  the Lorentz generator is of the form
\begin{equation}
\begin{array}{c}
 {\bf J}={\bf p}\times {\bf x} +
{\bf p}_+\times {\bf e}_+ +{\bf p}_1\times {\bf e}_1= \\
={\bf p}\times {\bf x} +c(2\epsilon -q){\bf e}_+ -2c{\bf e}_-.
\end{array}
\end{equation}
Hence, the Poincar\'e  invariants (Casimirs) read
\begin{equation}
\begin{array}{c}
{\bf p}^2/c^2=2qx -\pi^2,\\
 {\bf p}{\bf J}/c^2=(2\epsilon-q)q-2x.
\end{array}
\label{cas}\end{equation}
Therefore, the system under consideration has  internal degrees of freedom,
so that different classical solutions have the same  mass and spin.

Let us reduce the Hamiltonian system, substituting
(\ref{deco}) and (\ref{pdeco}) into (\ref{1})-(\ref{2}).
The resulting symplectic one-form reads
\begin{equation}
{\cal A}={\bf p}d({\bf x}+\frac{2{\bf e}_1}{q})+ \frac{2cd\pi}{q}+
\frac{{\bf pJ}}{cq}{\bf e}_+d{\bf e}_1,
\end{equation}
while the Lorentz generator is of the form
\begin{equation}
{\bf J}={\bf p}\times({\bf x}+\frac{2{\bf e}_1}{q})+
\frac{{\bf pJ}}{cq}{\bf e}_+\times{\bf e}_1.
 \end{equation}

Now  it is convenient to
fix the mass $m$ and spin $s$ of the system imposing
\begin{equation}
{\bf p}^2=m^2,\quad {\bf pJ}=ms,
\end{equation}
and  to introduce, instead of
${\bf e}_+,{\bf e}_1$, the new variables \begin{equation}
{\bf E}_{1}= {\bf e}_1+\frac{\pi{\bf e}_+}{q},\;\;
{\bf E}_2= \frac{m{\bf e}_+}{cq}-\frac{{\bf p}}{m},
\end{equation}
which obey the conditions
\begin{equation}
{\bf p}{\bf E}_a=0,\;\;{\bf E}_a{\bf E}_b=
-\delta_{ab},\;\;a=1,2.
\end{equation}
Then, introducing  the  complex coordinate
\begin{equation}
{\bf z}={\bf E}_1+i{\bf E}_2,
\end{equation}
one can represent the constraints in the conventional form
\begin{equation}
{\bf z}^2= 0, \quad {\bf z}{\bf{\bar z}} +1= 0,\quad {\bf p}{\bf z}= 0.
\label{const}\end{equation}
In these terms, the symplectic structure is of the form
\begin{equation}
\omega_{red}=d{\bf p}\wedge d{\bf X}+
isd{\bf z}\wedge d{\bf{\bar z}}+\frac{2cd\pi\wedge dq}{q^2},
\end{equation}
while the Lorentz generator reads
\begin{equation}
{\bf J}={\bf p}\times{\bf X}+
is{\bf z}\times{\bf {\bar z}},
\label{ajz}\end{equation}
where  we introduced the ``effective" coordinate
\begin{equation}
{\bf X}={\bf x}+\left(\frac{2}{q}+\frac{s}{m}\right)
{\bf e}_1+\frac{s\pi}{mq}{\bf e}_+.
\end{equation}
One may resolve the constraints  (\ref{const}) by noticing that
the first two of them imply that ${\bf z}$ may be written in
terms of a single complex
parameter $\omega$ as
\begin{equation}
{\bf z} = {\alpha\over \imath (\omega-\bar\omega)}
\left( 1+\omega^2 , 1 -\omega^2 , 2\omega \right).
\end{equation}
From the remaining constraint ${\bf pz}=0$ it  follows
that
\begin{equation}
\omega ={i p_2 \pm m\over p_0 + p_1}.
\end{equation}
So, one can finally write the symplectic structure
  and Lorentz
generator ${\bf J}$ solely in terms of ${\bf X}$, ${\bf p}$
and $q,\pi$.
It is easy to see that the reduced  symplectic structure reads
\begin{equation}
 d{\bf p}\wedge d{\bf X}\pm
s\frac{({\bf p}\times d{\bf p})\wedge d{\bf p}}{2m^3} +
\frac{2cd\pi\wedge dq}{q^2}.
\label{d}\end{equation}
The first two terms in (\ref{d}) define the symplectic structure
of  the standard ``minimal covariant model" for
anyons \cite{jakiw}, from  which  follows that the
spin $s$ is not quantized.

To analyze the "nonrelativistic" part of the system, one can reduce it by
${\bf p}$, and get the one- dimensional nonrelativistic mechanics
with a cubic potential
\begin{equation}
\pi\wedge dq,\quad
\pi^2 +q^3-2\epsilon q^2 +\frac{ms}{c^2}q + \frac{m^2}{c^2}=0
\end{equation}
Thus,  the spectrum of the system under consideration
contains both massive and tachionic branches, which have no
upper/lower bounds, respectively.
Nevertheless, this potential has a local minimum, where the
so-called ``semidiscrete" (``semistationary")
 or resonance-like  levels can exist,
 which are responsible for
numerous interesting phenomena \cite{qm}.
Notice that the local minimum of this system (``ground state")
corresponds to the  point $q=q_0$ defined by the equation
$$3q^2_0-4\epsilon q_0+\frac{ms}{c^2}=0,$$
where the mass and spin are related by the expression
$$\left(m^2+\frac{3\epsilon ms}{4c^2}+\frac{8\epsilon^3}{27}\right)^2
=\frac{4}{3}\left(\frac{ms}{c^2}-\frac{4\epsilon^2}{3}\right)^3.$$
For example, in the simplest case $\epsilon=0$, the ground
state is tachionic,
while  the ``mass" and spin are related as follows
$$ m=\frac{4s^3}{9c^6}.$$

Notice, that four-dimensional particle systems, defined by the Lagrangians
linear on torsion (third curvature),
are also related with one-dimensional non-relativistic
mechanics. For example, the symplest four- dimensional system of this sort,
formulated on non-isotropic curves, is connected by one-dimensional conformal
 mechanics \cite{cm}.\\

{\large  Acknowledgments.}
A.N. is thankful to E.Ramos, initiating his interest in the systems on
null-curves and V.Ter-Antonian for useful comments on the quantum mechanics
with a cubic potential.

\end{document}